\documentclass[pre,twocolumn,superscriptaddress,showpacs]{revtex4-1}

\usepackage{graphicx}  
\usepackage{dcolumn}  
\usepackage{amsmath}
\usepackage{amssymb}
\usepackage{bm}      
\usepackage{color}   
\usepackage{verbatim}
\usepackage[normalem]{ulem} 
\usepackage{times}
\usepackage{color}

\definecolor{yblue}{rgb}{0.06, 0.3, 0.57}
\usepackage[pdftex]{hyperref}
\hypersetup{colorlinks=true,linkcolor=yblue,citecolor=yblue,urlcolor=yblue}

\begin{document}

\title{Distribution of interevent avalanche times in disordered and
frustrated spin systems}

\author{John Ferre}
\affiliation{Physics Department, University of California, Davis,
California 95616, USA}

\author{Amin Barzegar}
\affiliation{Department of Physics and Astronomy, Texas A\&M University,
College Station, Texas 77843-4242, USA}

\author{Helmut G. Katzgraber}
\affiliation{Microsoft Quantum, Microsoft, Redmond, WA 98052, USA}
\affiliation{Department of Physics and Astronomy, Texas A\&M University,
College Station, Texas 77843-4242, USA}
\affiliation{Santa Fe Institute, 1399 Hyde Park Road, Santa Fe, New
Mexico 87501 USA}

\author{Richard Scalettar}
\affiliation{Physics Department, University of California, Davis,
California 95616, USA}

\begin{abstract}

Hysteresis loops and the associated avalanche statistics of spin
systems, such as the random-field Ising and Edwards-Anderson spin-glass
models, have been extensively studied.  A particular focus has been on
self-organized criticality, manifest in power-law distributions of
avalanche sizes. Considerably less work has been done on the statistics
of the times between avalanches. This paper considers this issue,
generalizing the work of Nampoothiri {\em et al.} [Phys.~Rev.~E {\bf
96}, 032107 (2017)] in one space dimension to higher space dimensions.
In addition to the interevent statistics of all avalanches, we also
consider what happens when events are restricted to those exceeding a
certain threshold size.  Doing so raises the possibility of altering the
definition of time to count the number of small events between the large
ones, which provides for an analog to the concept of natural time
introduced by the geophysics community with the goal of predicting
patterns in seismic events.  We analyze the distribution of time and
natural time intervals both in the case of models that include only
nearest-neighbor interactions, as well as models with (sparse)
long-range couplings.

\end{abstract}

\pacs{75.50.Lk, 75.40.Mg, 05.50.+q}

\maketitle

\section{\label{sec:intro}Introduction}

Many physical systems, when perturbed, respond in discrete jumps between
metastable states. The earth's tectonic plates provide an example of
such behavior in the form of earthquakes, which release a large amount
of energy before being pinned again \cite{klein:97}.  Similarly, a sheet
of paper creases and tears in jerky movements, resulting in crackling
sounds \cite{salminen:02}, the vortex lines of type-II superconductors
depin when the electric current becomes large enough \cite{goodman:66},
and the magnetic dipoles of ferromagnets align with a changing external
magnetic field in individual steps \cite{mayergoyz:86a,mayergoyz:91}.

In these situations, and many others, the system waits in its new
configuration until further changes in a driving field induce the next
jump. The history of the sample is of great importance.  The
configuration of the system is not just a function of the instantaneous
value of the drive but depends on the path followed.  In this
paper, we study this phenomenon from a relatively new perspective
which focuses on the distribution of interevent times and how that
distribution is affected by the introduction of a threshold in the
definition of an event. Our goal is not only to gain additional insight
into the detailed mechanism of hysteresis but also to examine the idea
of {\em natural time} \cite{rundle:16} in a more simple context than the
geophysical applications that have mainly been considered up to now.

Klein {\em et al.}~have suggested \cite{klein:97} that an alternate
approach to the prediction of large earthquakes is to use as a clock the
number of smaller earthquakes rather than quantifying intervals via a
traditional counting of days and years. Varotsos {\em et al.} first
introduced the term {\em natural time} to describe this procedure
\cite{varotsos:05,varotsos:11}. Recent investigations have studied this
concept in complex stochastic nonlinear processes, including its use in
characterizing the current state of a system as it progresses between
events \cite{rundle:17}. Investigating natural time with geophysical
data is difficult owing to the absence or incompleteness of historical,
and even modern, data on small earthquakes; large earthquakes are,
fortunately, not excessively common.  Additionally, controlled
experiments are out of the question.

Here we use numerical simulations to analyze hysteresis and natural time
in the context of several simple disordered and frustrated Ising spin
models
\cite{fisher:87,bray:86,mcmillan:84a,newman:92,parisi:79,mezard:87,binder:86}
exhibiting magnetic hysteresis: the three-dimensional random-field Ising
model, the Sherrington-Kirkpatrick model, and the Viana-Bray model. Our
key results are: (i) The distribution of interevent times between all
avalanches scales with the number of lattice sites for the random-field
Ising model and the Viana-Bray model but not for the
Sherrington-Kirkpatrick model.  (ii)  The pseudogap exponent $\theta$,
which characterizes the behavior of the interevent distribution for
vanishing interevent time, is zero.  (iii) The addition of long-range
interactions decreases the number of small interevent times but does
not affect the statistics of the intervals between large events, nor do
they alter $\theta$. (iv) Despite exploring various models and parameter
regimes, we fail to find a situation where the predictive capability of
the natural time method is strong for spin avalanches in magnetic
hysteresis. (v) By imposing a minimum avalanche size threshold,
different models can be classified by their interevent distribution.
(vi) Finally, at a sufficiently large minimum avalanche size threshold,
the interevent time in the Sherrington-Kirkpatrick model follows a
Weibull distribution with shape factor $k\sim 1$, i.e., a Poisson
distribution.

The use of simulations allows us to generalize to higher dimensions a
recent analytical study by Nampoothiri {\em et
al.}~\cite{nampoothiri:17} on the interevent time distribution of the
one-dimensional random-field Ising model.  The central result of that
work was the computation of the distribution of times $P(\Delta B)$ of
the magnetic field change $\Delta B$ between spin avalanches. (If the
magnetic field is increased at constant rate, $\Delta B$ is proportional
to time.) It was found that $P(\Delta B) \sim (\Delta B)^\theta$ as
$\Delta B \rightarrow 0$ \cite{nampoothiri:17}, with $\theta=0$ for the
short-range ferromagnetic random-field Ising model, whereas $\theta =
0.95$ for the long-range antiferromagnetic case. Other studies of the
distribution of gaps between events have been conducted in the context
of amorphous solids and hard frictionless spheres
\cite{lin:14,kamakar:10,mueller:15,lin:14b,wyart:12}. There, when the
strain $\gamma$ is sufficiently increased, a corresponding stress drop
follows. The distribution of gaps $\Delta \gamma$ highlights differences
between the yielding and depinning processes and reveals information on
mechanical stability.

The paper is structured as follows. Section \ref{sec:model} introduces
the models studied, as well as the algorithmic approach and analysis
methods used. In Sec.~\ref{sec:results1} we present results on the
statistics of interevent times for all avalanches, and in
Sec.~\ref{sec:results2} we repeat the analysis with the introduction of
an event threshold. Section \ref{sec:results3} discusses the effects of
additional small-world bonds between the variables, followed by a study
of return point memory and concluding remarks in
Sec.~\ref{sec:conclusions}.

\begin{figure}[t]
\centering
\includegraphics[width = \columnwidth]{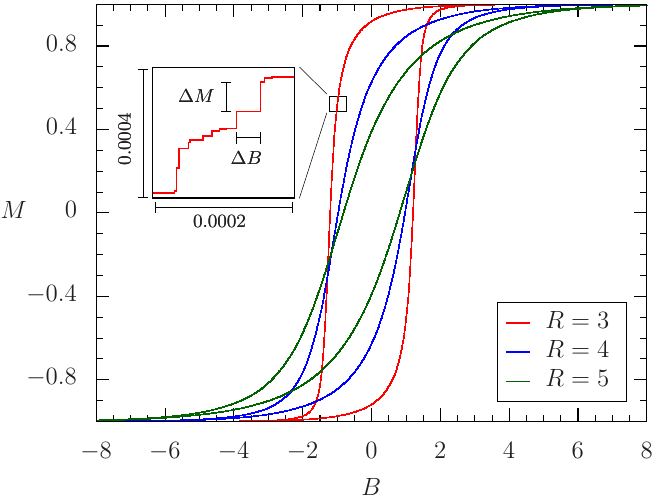} 
\caption{
Hysteresis loop of the random-field Ising model of size $N=100^3$ with a
distribution of fields of widths $R = 3$, $4$, and $5$. On a large
scale, the curves appears to represent a smooth evolution of the
magnetization $M(B)$.  However, as can be seen in the inset, the
magnetization $M(B)$ is composed of a series of discrete jumps. $\Delta
M$ is the change in magnetization. $\Delta B$ is the interevent time.
} 
\label{fig:hysteresis} 
\end{figure}

\section{\label{sec:model}Model and Methods}

We first consider the random-field Ising model (RFIM) defined by the
Hamiltonian
\begin{align}
{\cal H}_{\rm RFIM}  = -J \sum_{\langle ij \rangle} S_i S_j 
- \sum_i h_i S_i - B \sum_i S_i .
\label{eq:rfim}
\end{align}
Here $S_i = \pm 1$ is a discrete degree of freedom at site $i$ of a
cubic lattice with $N$ sites, and $\langle ij \rangle$ represents a sum
over nearest neighbors.  $J$ is the exchange constant between
nearest-neighbor sites, and units are set so that $J=1$. Each site $i$
is assigned a random magnetic field $h_i$, drawn from a Gaussian
distribution
\begin{equation}
P(h_i) = \frac{1}{\sqrt{\pi} R}\, {\rm exp}(-h_i^2/R^2).
\end{equation}
$R$ controls the width of the distribution and thereby the strength of
the disorder. A spatially uniform field $B$ is used to drive the
hysteresis loop.

Unlike the ferromagnetic Ising model where $h_i = 0$ $\forall i$, the
RFIM does not exhibit ferromagnetic order in $d=2$ (or below). As the
temperature is lowered in higher space dimensions, however, a freezing
transition occurs.  This is followed by a ferromagnetic transition
\cite{nattermann:88,belanger:88,belanger:91,rieger:95a,nattermann:98}.
Here we are not concerned with these aspects of the equilibrium
finite-temperature phase diagram but instead focus on the evolution of
the magnetization as the external field $B$ is sequentially changed,
with dynamics defined by each spin remaining parallel to its local
environment at each step, that is, effectively $T=0$ (see below).

\begin{figure}[t] 
\centering \includegraphics[width =\columnwidth]{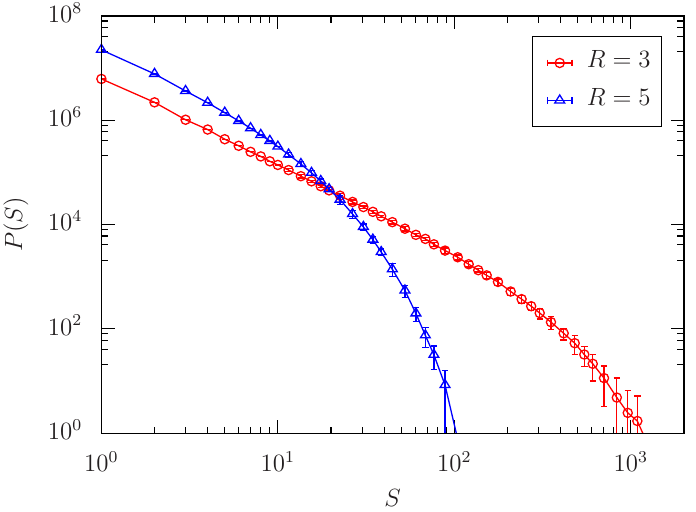} 
\caption{
Distribution of avalanche sizes for field width $R = 3$ and $5$. The
distribution is known to have power-law behavior for $R=R_{c} \approx
2.16$.  A critical region, for which power law behavior persists for
several decades of avalanche size, occurs up to $R \approx 4$. For 
details see Ref.~\cite{perkovic:95}.}
\label{fig:avaSize} 
\end{figure}

Typical RFIM hysteresis curves are shown in Fig.~\ref{fig:hysteresis}.
From them, we can extract the size of all the individual magnetization
jumps $S$ and hence their distribution. It is well known
\cite{perkovic:95} that this distribution has power-law behavior at low
disorder, which we reproduce in Fig.~\ref{fig:avaSize}. We can also
extract the distribution of time (as measured by the change in external
field $\Delta B$) between events.  The latter quantity has been much
less studied than the former.

The Sherrington-Kirkpatrick model (SKM) \cite{sherrington:75} is given by
the Hamiltonian
\begin{equation}
{\cal H}_{\rm SKM} = -\sum_{i<j}{J_{ij}S_iS_j} - B \sum_i S_i . 
\end{equation}
In the SKM, every site $i \in \{1, \ldots, N\}$ interacts with every
other site $j$ via $J_{ij}$.  That is, the interaction is infinite
range.  The exchange constants $J_{ij}$ are disordered, and in our study
are given by a Gaussian distribution with zero mean and a standard
deviation of $J_0$.  We set $J_0=1$ as our unit of energy. The SKM shows
self-organized criticality (SOC) for all disorder \cite{andresen:13,sharma:18}.

Finally, we also study the Viana-Bray model (VBM) \cite{viana:85} in
which each spin is randomly connected to $z=6$ other spins.  Thus, the
VBM is still long ranged but with a finite coordination number. Unlike the SKM, the VBM
does not have SOC \cite{andresen:13}. Note that the RFIM has an
explicit parameter $R$ with which the disorder strength can be tuned
whereas the SKM and VBM do not.

In order to generate a hysteresis loop for the RFIM, we compute 
the local fields,
\begin{align} 
B_i = -h_i - J \sum_{j \in {\cal N}(i)} S_j\,, 
\label{eq:localRFIM} 
\end{align}
and for the SKM and VBM,
\begin{align} 
B_i = - \sum_{j \in {\cal N}(i)}J_{ij} S_j\,. 
\label{eq:localSKVB} 
\end{align}
For the RFIM, ${\cal N}(i)$ includes the nearest neighbor spins, whereas
${\cal N}(i)$ for the SKM consists of all spins, while the six randomly
chosen neighbors define ${\cal N}(i)$ for the VBM.

A hysteresis loop is generated as follows: Starting at $B=\infty$ we
reduce $B$ to a value $B_k={\rm max}\{B_i\}$. This is the external field
at which the spin $S_k$ becomes unstable. $S_k$ is then reversed and the
reconfiguration of the lattice and of the collection of local fields
$\{B_i\}$ is computed based on greedy dynamics \cite{schnabel:17}. Once
$S_k$ flips and the local fields are recomputed, the next most unstable
spin $l$ is flipped, i.e.,~its (updated) local field is now greater than
the external field: $B_l > S_l \cdot B$. This process is continued until all
unstable spins are reversed. The total count of spins flipped is
recorded as the size of the associated avalanche. The avalanche size
determines the change in magnetization $\Delta M$, which is twice the
total fraction of spins which flip. At this point, the external field $B$
is reduced once again to the next largest $\{B_i\}$ and the process is
repeated until all spins are flipped and the system reaches saturation
but with the opposite sign of the magnetization.  The interevent times
are the values $\Delta B$ that the external field jumps between each
completed avalanche.

We begin, in Sec.~\ref{sec:results1}, by analyzing the distribution of
time intervals $P(\Delta B)$ which results from using the most broad
definition of an avalanche, i.e., by including even the smallest
possible $\Delta M=2/N$, resulting from a single spin flip. We also
calculate the pseudogap exponent $\theta$, given by $P(\Delta B) \sim
(\Delta B)^\theta$ as $\Delta B \rightarrow 0$. This follows the
procedure described in recent literature \cite{nampoothiri:17} on the
one-dimensional RFIM.

Next, in Sec.~\ref{sec:results2} we use a minimal threshold $\Delta M$,
only above which a change in a spin configuration is considered an
event. We analyze how the distribution in interevent times is affected
by making $\Delta M>2/N$. The introduction of such a threshold $\Delta
M$ allows us to consider alternate measures of the {\em interval}
between events. Specifically, we can define the {\em natural time}
$\Delta A$ between large avalanches by counting the number of small
avalanches (those with $\Delta M$ less than the threshold) which occur
between large ones. We also define the {\em total natural time} $\Delta
F$ between large avalanches to be the total number of flipped spins
(i.e., the change in magnetization) which has accumulated.  This latter
procedure weights each small avalanche by the number of spins which
turned over.

\begin{figure}[t]
\centering
\includegraphics[width =\columnwidth]{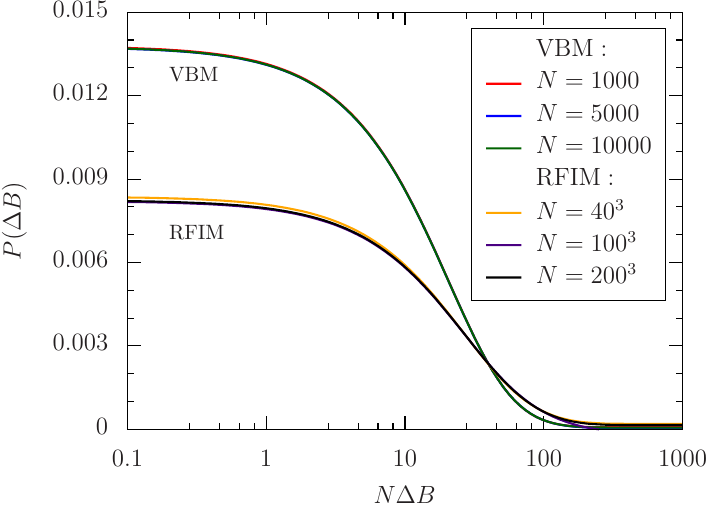}
\caption{
Scaling collapse of the distribution of interevent times for the RFIM
at field width $R=2.3$ and the VBM for various system sizes $N$. If the
field interval axis is scaled by $N$, the distributions for different
$N$ coincide.  That is, $P(N,\Delta B) \sim \tilde P(N \Delta B)$. The
vanishing slope at small $\Delta B$ indicates the pseudogap exponent
defined by $P(\Delta B) \sim (\Delta B)^\theta$ as $\Delta B \rightarrow
0$ obeys $\theta = 0$, in agreement with analytic results in one space
dimension ($d=1$).
}
\label{fig:deltaH}
\end{figure}

These different approaches to the interevent time are chosen to parallel
analogous definitions in the geophysics community \cite{varotsos:06} where
natural time employs only earthquakes exceeding a certain size as
events and the number of small earthquakes between the large
ones is recorded. The current state of an earthquake cycle is analyzed
by constructing a cumulative distribution function of interevent times
between large earthquakes, which shows a Weibull form \cite{rundle:16},  
\begin{equation}
\label{eq:Weibull}
f(t) = 1 - e^{-(t / \lambda)^k}.
\end{equation}
Here $\lambda$ is the scale parameter and $k$ is the shape parameter.
For $k<1$ the cumulative probability function of the Weibull
distribution has an initial rapid rise, while if $k>1$ the initial slope
is small. If $k=1$, Eq.~\eqref{eq:Weibull} becomes the interevent
distribution of a Poisson process.  The Weibull distribution is commonly
used in the materials science community to characterize the time to
failure, where $k<1$ corresponds to a failure rate which decreases with
time. In contrast, $k>1$ corresponds to a failure rate which increases
with time.  Motivated by the geophysics problem, we perform a similar
fit to the cumulative distribution of interevent times in our spin
model hysteresis loops. By taking the natural logarithm of both sides
twice, Eq.~\eqref{eq:Weibull} becomes
\begin{equation}
{\log}[-{\log}(1-f(t))]=k{\log}(t)-k{\log}(\lambda),
\label{eq:loglog}
\end{equation}
so that a plot of the data in the form ${\log}[-{\log}(1-f(t))]$
vs ${\log}(t)$ yields a linear relation if $f(t)$ has a Weibull form.

\section{\label{sec:results1}Statistics of All Avalanche interevent Times}

We start by analyzing the distributions of time intervals that occur
between every avalanche, including avalanches of a single flip, i.e.,
$\Delta M =2/N$. When an avalanche occurs, we mark its magnetic field
value. Then, we can define $\Delta B$ as the difference between any two
consecutive avalanches and accumulate the distribution of interevent
times as $P(\Delta B)$.

In order to compare distributions of different parameters properly,
$P(\Delta B)$ is normalized. This is done by dividing by the total
number of intervals $N_{\rm int}$. As $R$ increases to large values, the
spins feel a wide range of random fields, and their local fields $B_i$
[Eq.~\eqref{eq:localRFIM}] become widely separated. In the limit $R
\rightarrow \infty$, all events become single flips because the
contribution to $B_i$ from the exchange interactions $J$ is negligible
in comparison.  The total number of intervals $N_{\rm int}$ approaches
the number of lattice sites (spins) $N$. Similarly, as $R$ decreases,
the total number of intervals becomes small. At $R=0$, the hysteresis
loop becomes completely square, $N_i=1$, and the entire lattice flips
from up to down at the single external field value $B=-2dJ$, where $d =
3$ is the space dimension. Normalizing $P(\Delta B)$ to $N_{\rm int}$
eliminates this trivial effect. The sum of all avalanches in the
distribution equals unity, independent of the choice of parameters.

\begin{figure}
\centering
\includegraphics[width = \columnwidth]{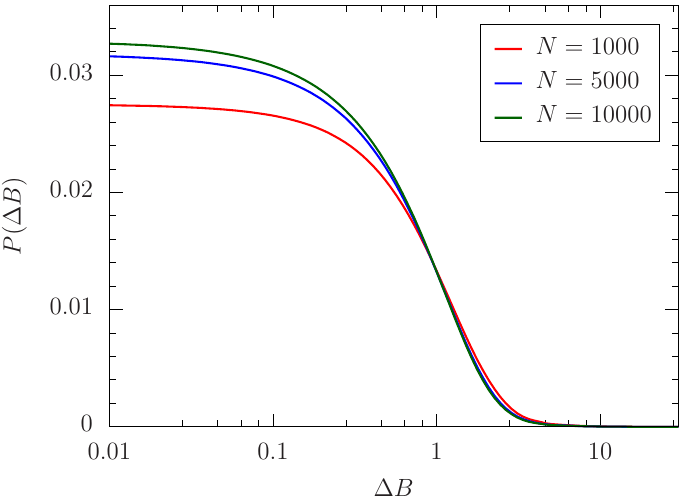}
\caption{
Distribution of interevent times for the SKM. The distributions largely
overlap, apart from a modest separation at small $\Delta B$. In sharp
contrast to the RFIM and VBM data of Fig.~\ref{fig:deltaH}, $P(N,\Delta
B)$ does not show a scaling collapse to $P(N\Delta B)$.
}
\label{fig:SKdist}
\end{figure}

We begin by analyzing the RFIM, plotting the distribution of interevent
times in Fig.~\ref{fig:deltaH}.  Distributions of varying lattice sizes
$N$ are seen to collapse if the event intervals $\Delta B$ are scaled by
$N$, that is
\begin{equation}
\label{eq:Pscaling}
P(N,\Delta B) \sim \tilde P(N \Delta B) \,.
\end{equation}
The interevent distribution for the VBM is similar to the RFIM (see
Fig.~\ref{fig:deltaH}) and scales with the number of variables $N$.
Due to
the fundamental difference between finite and diverging number of
neighbors
\cite{andresen:13}, we expect the same scaling behavior for
differing finite coordination numbers of the VBM.

As can be seen in Fig.~\ref{fig:SKdist}, the SKM distributions for
different lattice sizes collapse with an {\it unscaled} $\Delta B$.
Other than that, the shape of the interevent distribution of all
avalanches is similar to the RFIM (Fig.~\ref{fig:deltaH}).  Because the
VBM is long ranged and the RFIM is short ranged, but both have similar
scaling forms for $P(\Delta B)$, we conclude that the connections
between distant spins do not by themselves give rise to a change from
the RFIM collapse with $N\Delta B$. Instead, the most likely cause of
scaling differences is the presence of SOC at all disorder strengths in
the SKM due to the fully-connected topology.

\begin{figure}
\centering
\includegraphics[width =\columnwidth]{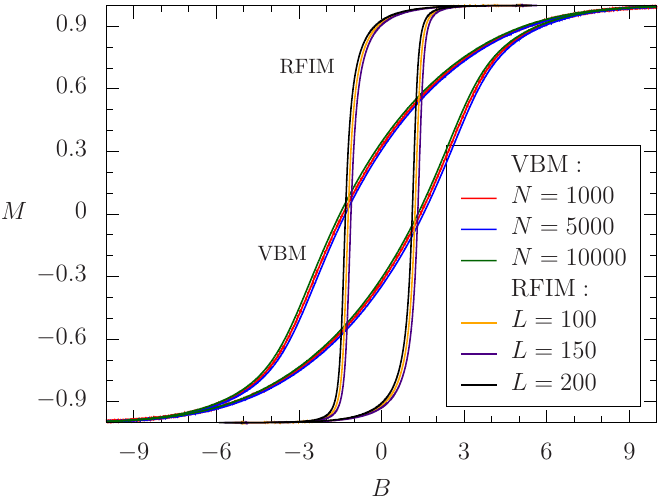}
\caption{
Hysteresis loop for the RFIM with $R=3$ and for the VBM for different
system sizes $N$. The shape is similar for various lattice sizes for all
disorders. As argued in the text, this independence of lattice size
underlies the scaling form $P(N,\Delta B) \sim \tilde P(N\Delta B)$.
}
\label{fig:RFhys}
\end{figure}

Another way to gain insight into the scaling of interevent times is to
observe the behavior of the hysteresis loops. 
While hysteresis loops have been extensively studied \cite{pazmandi:99, andresen:13, zhu:13, sharma:14, sarflyer:12}, we focus on the width of the loop in relation to the lattice size.
For the RFIM and VBM, the
loops are the same width across all lattice sizes (see
Fig.~\ref{fig:RFhys}). This means that the total time $T$ for traversal
of the loop is constant. In the limit where events are small and fairly
isolated spatially, the number of events grows linearly with lattice
size $N$, and the time $\Delta B$ between events is proportional to
$1/N$. This picture offers a qualitative explanation of the dependence
of $P$ on $N \Delta B$.

\begin{figure}
\centering
\includegraphics[width=\columnwidth]{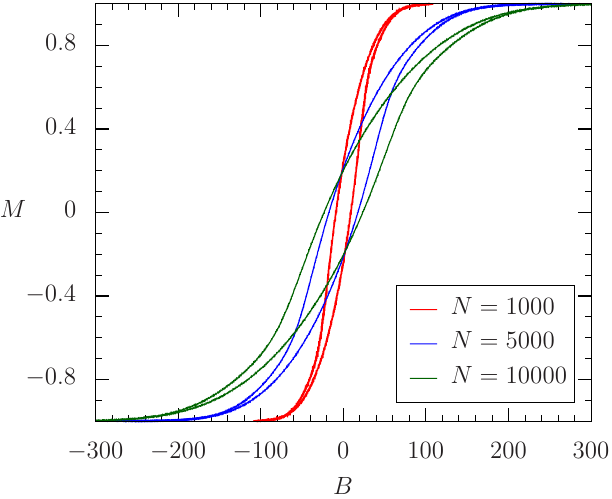}
\caption{
Hysteresis loop for the SKM. The width of the loop (e.g., at $M=0$)
increases with lattice size: $\Delta B \sim 9$, $17$, and $23$ for
$N=1000$, $5000$, and $10000$, respectively. As argued in the text, this
dependence on lattice size underlies the difference in scaling behavior
from the RFIM and VBM cases.}
\label{fig:SKhys}
\end{figure}

The hysteresis loop for the SKM (Fig.~\ref{fig:SKhys})  is different
from the VBM and the RFIM. Its width grows as the lattice size is
increased -- the total time across the loop increases with $N$. If we
again consider a limit where events are small and fairly isolated
spatially, so that the number of events grows linearly with lattice size
$N$, the interval between individual events $\Delta B$ is expected to be
roughly independent of lattice size.

\begin{figure}[t]
\centering
\includegraphics[width=\columnwidth]{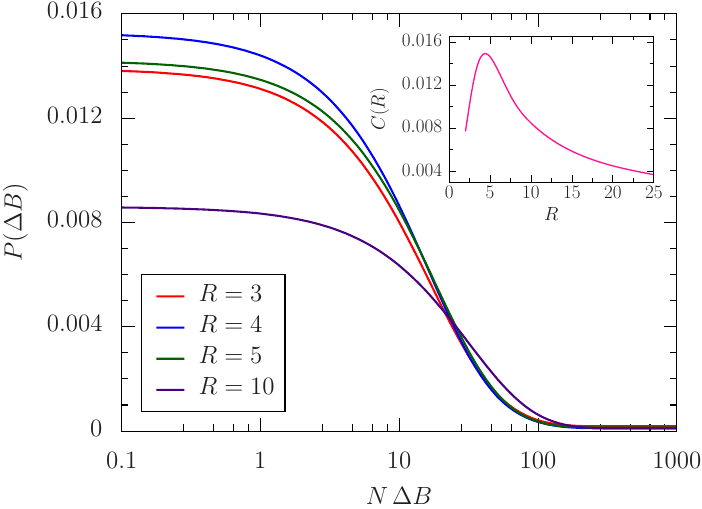}
\caption{
Distribution of interevent times for the RFIM for various field widths
$R$. $C(R)$ [the value of the interevent distribution at small $\Delta
B$, Eq.~\eqref{eq:CR}] is highest for $R = 4$. The behavior of $C(R)$ is
nonmonotonic. The slope of $P(\Delta B)$ is zero for small $\Delta B$,
confirming that $\theta = 0$ for all $R$. Inset: $C(R)$
[Eq.~\eqref{eq:CR}] for a fine mesh of $R$ values. This allows for a more refined determination of the
peak position, $R \approx 3.7$. This is in qualitative agreement with the $d=1$
analytical result, which has a similar peak at $R = 1$
\cite{nampoothiri:17}.
}
\label{fig:deltaHmulti}
\end{figure}

Returning to the RFIM, we consider the dependence of the limit of
interevent time distribution at small $\Delta B$ on disorder width $R$.
Figure \ref{fig:deltaHmulti} shows the distributions for different $R$
values.  The quantity
\begin{equation}
\label{eq:CR}
C(R) = \lim_{N\Delta B \rightarrow 0} \tilde P(N\Delta B)
\end{equation}
characterizes the value of the distribution at the smallest interval
sizes. The inset to Fig.~\ref{fig:deltaHmulti}, showing $C(R)$, exhibits
a peak for $R \approx 3.7$. Note that for the analytical calculation of
the one-dimensional case, $C(R)$ peaks at $R \approx 1$
\cite{nampoothiri:17}.

The flatness of $P(\Delta B)$ as $\Delta B \rightarrow 0$ implies that
the pseudogap exponent $\theta=0$ for all $R$ in the RFIM. It likewise
vanishes for the SKM and VBM. This value for the exponent is the same as
the one-dimensional RFIM \cite{nampoothiri:17}.  It has been argued that
this is a consequence of the mapping between the RFIM and a depinning
process when the dimensionality is less than $5$ \cite{bruinsma:84}. The
depinning process is known to have $\theta = 0$ \cite{lin:14}.  Thus our
results confirm previous conjectures on the nature of the gap
statistics.

The interevent time distributions $P(\Delta B)$ of
Figs.~\ref{fig:deltaH} and \ref{fig:SKdist} illustrate the
unpredictability of avalanche occurrences.  The distributions have
significant weight over several orders of magnitude of $\Delta B$. The
time between avalanches, as measured by the traditional definition
$\Delta B$ (which is proportional to the conventional time interval
$\Delta t$ if $B$ is swept at constant rate), takes a very wide range of
values. In Sec.~\ref{sec:results2} we explore whether alternate
definitions might yield a narrower distribution and hence more
predictable intervals.

\section{\label{sec:results2}Statistics of Above Threshold Interevent
Times}

In Sec.~\ref{sec:results1} we have seen that the distribution of
interevent times $P(\Delta B)$ is very broad when all avalanches are
considered as events. We now reanalyze the distribution but impose an
event threshold $L_A$. This both eliminates the small (and therefore
presumably more random) magnetization jumps and opens the door to
counting the number of jumps as an alternate definition of interevent
time. This latter procedure follows suggestions in the geophysics
community where including an avalanche threshold was argued to help
determine where a certain geographic region is located in the earthquake
cycle.

In geophysics studies, the imposition of a threshold was shown to lead
to the interevent distribution following a Weibull process
\cite{rundle:16}.  Here we use a similar approach, and verify if the
statistics obey the same distribution.  We examine several definitions
of the interevent time: $\Delta A$ characterizes the number of small
avalanches, $\Delta F$ the total number of individual flips.  This
complements the use of $\Delta B$, the change in magnetic field between
events (see Sec.~\ref{sec:results1}).

\begin{figure}
\centering
\includegraphics[width =\columnwidth]{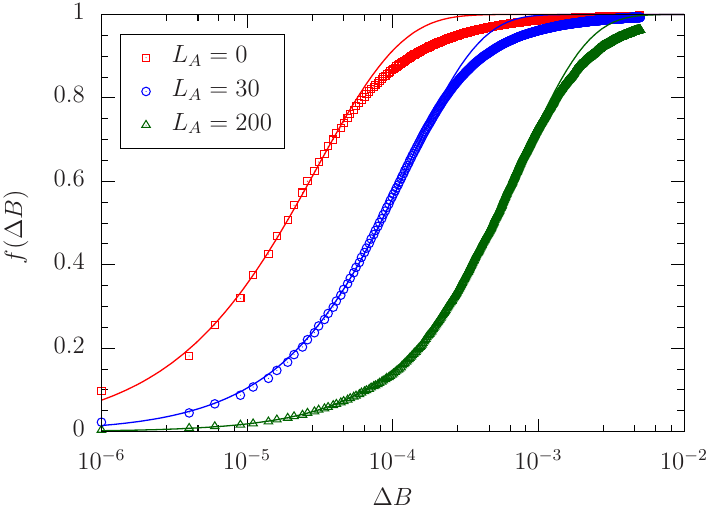}
\caption{
Cumulative distribution of the interevent times for the RFIM at $R =
3$. The lattice size is $N= 100^3$. The data (points) are plotted along
with the Weibull fit (lines). Several large avalanche thresholds are
imposed. The Weibull cumulative distribution function appears to fit
reasonably well to the cumulative distribution of the data.  This occurs
for $\Delta A$ and $\Delta F$ as well. The deviations are examined more
critically in Fig.~\ref{fig:loglogBR}, which better emphasizes
deviations from the Weibull form.}
\label{fig:weibullB}
\end{figure}

If large avalanches tended to occur after relatively {\it constant}
numbers of small avalanches $\Delta A$, then a plot of the cumulative
distribution function $f(\Delta A)$ would take the form of an abrupt
step, reflecting a sharply peaked probability $P(\Delta A)$, i.e., large
events separated by {\it one specific} $\Delta A$.  Figure
\ref{fig:weibullB} shows the cumulative distribution functions for
different disorder strengths $R$ and for different choices of the
threshold $L_A$ for counting small avalanches.  We see no significant
tendency for the cumulative distribution to become more steplike than
when plotted as a function of $\Delta B$.

Although the continued broadness of the distributions of
Fig.~\ref{fig:weibullB}---despite the replacement of $\Delta B$ by
$\Delta A$ and $\Delta F$---suggests that natural time does not sharpen
the distribution of interevent spacing, we can still ask whether the
underlying distributions of interevent times are similar to those found
in geophysics applications. Figure \ref{fig:weibullB} shows the Weibull
fits to the distributions $f$ in addition to the raw data.

While naively it appears that the data of Fig.~\ref{fig:weibullB} might
follow a Weibull interevent distribution, a more discerning check is
made by plotting the data with modified axes:
${\log}[-{\log}(1-f(t))]$ vs ${\log}(t)$. On these axes, the data
should form a straight line, as discussed earlier [see
Eq.~\eqref{eq:loglog}].

\begin{figure}[t!]
\centering
\includegraphics[width =\columnwidth]{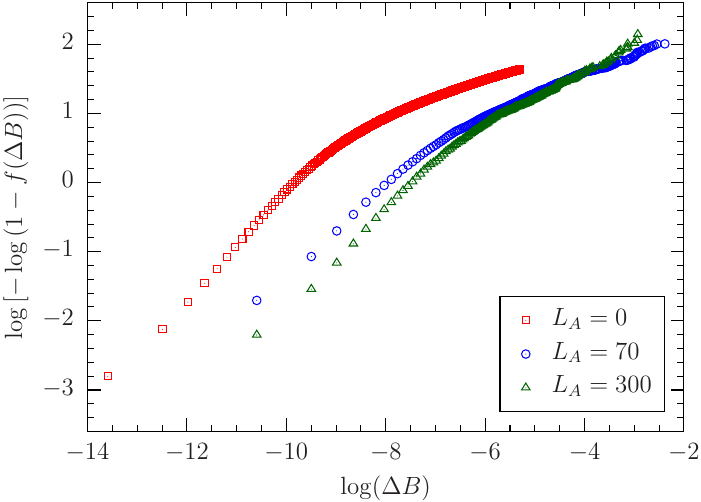}
\caption{
Cumulative distribution of the interevent times for the RFIM at $R=2.3$
on a double-logarithmic scale. For the distribution to be fit by a
Weibull distribution, the data must be linear. Instead, the curves show
a significant downward concavity for all values of $L_A$. The same trend
is found for $\Delta A$ (not shown).  }
\label{fig:loglogBR}
\end{figure}

\begin{figure}[t!]
\centering
\includegraphics[width =\columnwidth]{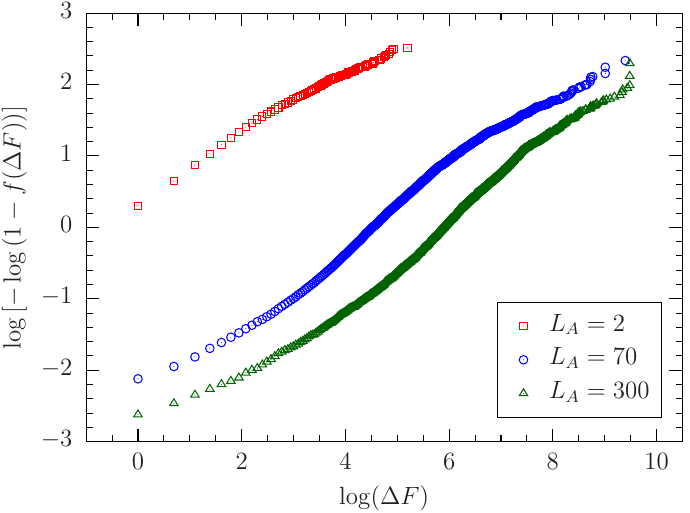}
\caption{
Cumulative distribution of the interevent times for the RFIM at $R=2.3$
on a double-logarithmic scale.  The data are closest to linear for the
smallest threshold.}
\label{fig:loglogFR}
\end{figure}

\begin{figure}[t!]
\centering
\includegraphics[width =\columnwidth]{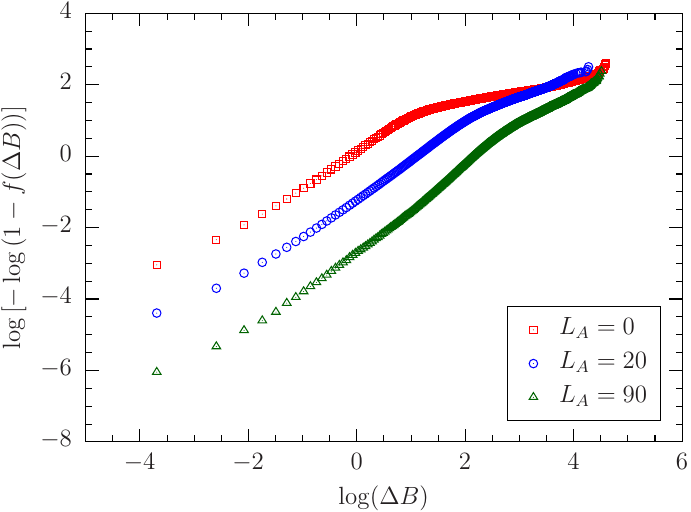}
\caption{
Cumulative distribution of the interevent times for the SKM with
lattice size $N=10000$. As $L_A$ is increased, the data become more
linear, which shows that the distribution of $\Delta B$ can be
approximated as Weibull at sufficiently large $L_A$. The slope for
$L_A=90$ is $k\sim 1$, which implies that the Weibull distribution
simplifies into a Poisson distribution. The same phenomenon occurs in
the VBM with a slope of $k\sim 0.8$.
}
\label{fig:loglogBSK}
\end{figure}

\begin{figure}[t!]
\centering
\includegraphics[width =\columnwidth]{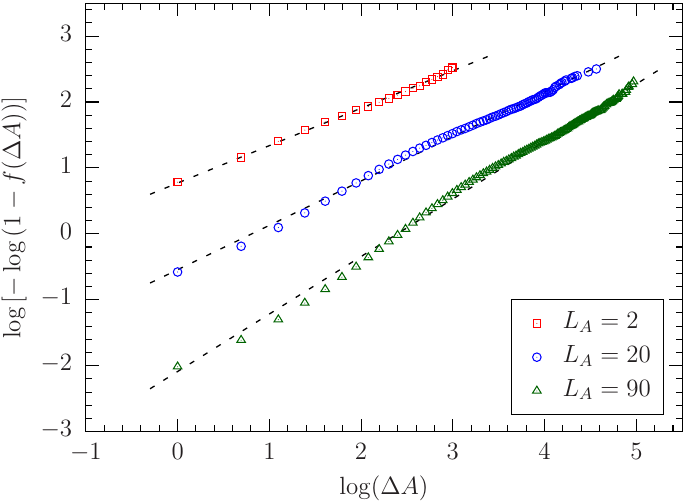}
\caption{
Cumulative distribution of the interevent times for the SKM with
lattice size $N=10000$. The data are approximately linear for all $L_A$,
i.e., the distribution is Weibull.
}
\label{fig:loglogASK}
\end{figure}

First, we analyze the RFIM, where the distributions do not appear to be
well fit by a Weibull distribution by any definition of time (see
Figs.~\ref{fig:loglogBR} and \ref{fig:loglogFR}). We focus on $R >
R_{\rm c} \sim 2.16$, the critical value of $R$ below which an infinite
avalanche occurs in which a macroscopic fraction of the spins all flip
at once \cite{perkovic:95}. Above $R_{\rm c}$, $\Delta A$ and $\Delta B$
yield a curve that is concave down, while $\Delta F$ yields a curve with
an inflection point. The fact that the distribution is concave down
implies that there is a scarcity of large-time intervals for the
distributions to be Weibull. The quality of the fits of RFIM interevent
times $\Delta B$ to a Weibull distribution (Fig.~\ref{fig:loglogBR})
does not appear to be very sensitive to the value of the avalanche
threshold.  In the case of natural time $\Delta F$
(Fig.~\ref{fig:loglogFR}) smaller threshold gives a somewhat better fit.
In either case, as $L_A$ increases, the average time between events
increases, which leads to a lower intercept with the vertical axis; this
follows from Eq.~\eqref{eq:loglog}. The situation is rather different
for the SKM when using the interevent time $\Delta B$. As $L_A$ is
increased, the Weibull fit improves significantly, as shown in
Fig.~\ref{fig:loglogBSK}.  Interestingly, the shape parameter $k$ is
close to unity, i.e.,~the distribution is Poissonian. The same fit
improvement with larger $L_A$ occurs for the VBM (not shown), but $k
\sim 0.8$ in that case. 
We are unable to determine if this value of $k$ varies with the VBM's coordination number $z$, and we might perform simulations of varying $z$ in a future paper.
The SKM also provides interesting results for
the natural time methods. $f(\Delta A)$ is rather close to a Weibull
distribution for all $L_A$ (Fig.~\ref{fig:loglogASK}). $\Delta F$,
however, provides a fit that worsens as $L_A$ is increased. The VBM
shows similar results in that both methods of natural time provide the
best fit at low $L_A$.

\section{\label{sec:results3}Effect of Longer Range Couplings}

On large lattices, small avalanches amongst clusters of spins which are
far from each other are likely to occur in a rather independent manner.
This might be problematic for periodic large-event intervals, because
avalanches which are decoupled are unlikely to provide a predictive
countdown to an above-threshold event.  In order to introduce a more
collective behavior of the entire cubic lattice, we introduce long-range
couplings by dividing the entire cubic lattice into randomly selected
pairs of sites. At the algorithmic level, this is accomplished by
starting with site $0$ and then randomly selecting one of the other
sites $p_0$ of the lattice as a partner to site $0$. Note that $p_0$ is
not allowed to be one of the existing six nearest neighbor sites. After
this is done, both sites $0$ and $p_0$ are eliminated as potential
partners and one proceeds to site $1$ (assuming $p_0 \neq 1$) and
randomly assigns it a partner $p_1$. This process is continued until all
sites in the cubic lattice have a seventh neighbor. When assigned in
this way, the probability of a site being part of a pair is independent
of the geometric proximity between the two sites (as long as they are
not nearest neighbors).  These longer-range neighbors are coupled by an
exchange constant $J'$, so that in the computation of the local field,
and hence the determination of whether to flip $S_i$,
Eq.~\eqref{eq:localRFIM} is generalized to include $p_i$ as part of
${\cal N}(i)$. Setting $J'=0$ recovers the original nearest neighbor
only model, and increasing $J'$ allows us, in a smooth manner, to
increase the long-range interactions across the lattice. Our model
Hamiltonian thus becomes
\begin{equation}
{\cal H}  = -J \sum_{\langle ij \rangle} S_i S_j - J'\sum_{i}S_i S_{p_i} 
- \sum_i h_i S_i - B \sum_i S_i , 
\label{eq:lr}
\end{equation}
where $p_i$ is the long-range site connected to site $i$.  A similar
procedure has previously been introduced in Ref.~\cite{scalettar:91a} to
study finite-temperature phase transitions in Ising models with
long-range interactions (which prove to be of mean-field character) and,
more generally, are considered in the context of small-world networks
\cite{albert:02}.

\begin{figure}[t]
\centering
\includegraphics[width =\columnwidth]{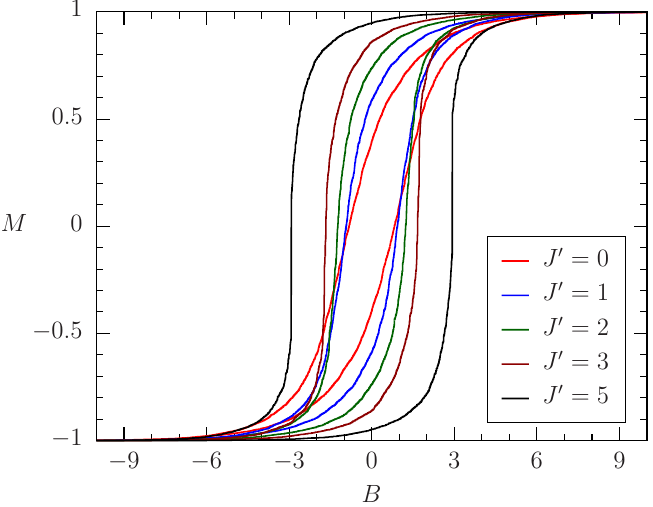}
\caption{
Hysteresis loops for the Hamiltonian in Eq.~\eqref{eq:lr} for various
$J'$ values. Here, $R = 5$, and the lattice size is $N = 30^3$. As the
strength of $J'$ is increased, the hysteresis loop becomes steeper and
wider. This is due to the increased correlation of lattice sites, which
causes avalanches to occur at the same magnetic field value.}
\label{fig:hysJs}
\end{figure}

\begin{figure}
\centering
\includegraphics[width =\columnwidth]{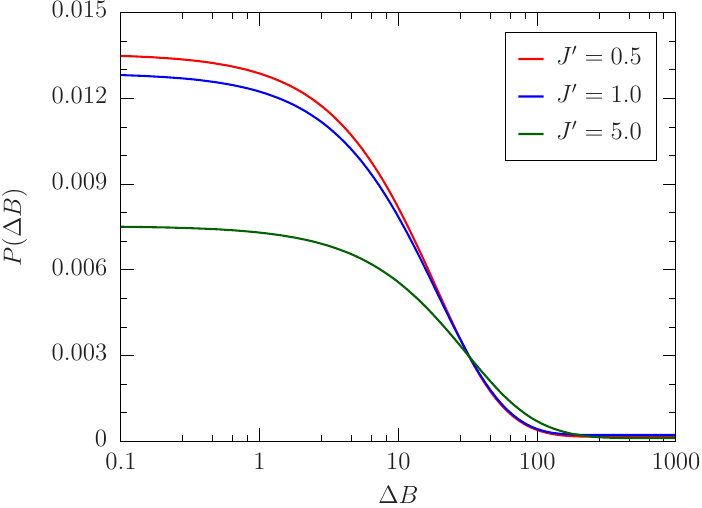}
\caption{
Distribution of interevent times for various values of $J'$.  The
distribution is always monotonic, and there are fewer small interevent
times when $J'$ is increased. This is due to the additional correlation
between avalanches.
}
\label{fig:PDistribution}
\end{figure}

Figure \ref{fig:hysJs} shows hysteresis loops for the Hamiltonian
presented in Eq.~\eqref{eq:lr}. The loops become steeper and the width
is increased with $J'$. This occurs because different lattice sites
become correlated, which causes avalanches to combine. The distribution
of interevent times $\Delta B$ is shown in
Fig.~\ref{fig:PDistribution}. For all $J'$ values, the distribution is
monotonically decreasing and larger $J'$ values suppress the frequency
of small avalanches. The pseudogap exponent is zero for any strength of
the long-range connections.

\begin{figure}
\centering
\includegraphics[width =\columnwidth]{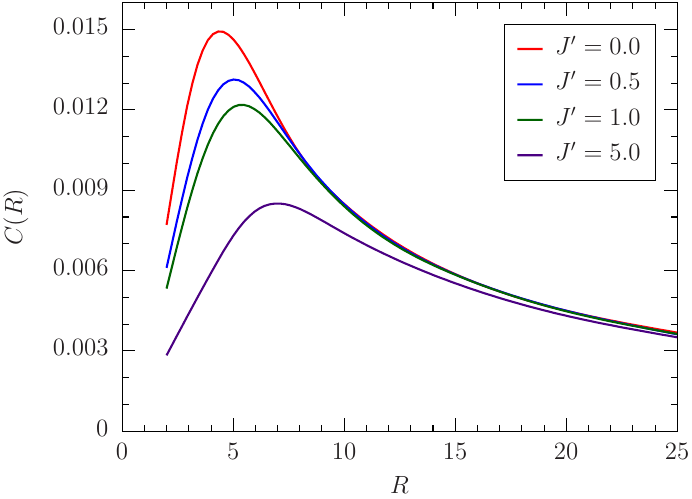}
\caption{
$C(R)$ [Eq.~\eqref{eq:CR}] plotted against the disorder strength $R$ for
various $J'$ values. $C(R)$ is lowered as $J'$ increases, and the curve
collapses onto the $J' = 0$ case when $R$ becomes large enough.}
\label{fig:CRallJs}
\end{figure}

$C(R)$, given by Eq.~\eqref{eq:CR}, is shown in Fig.~\ref{fig:CRallJs}
for different values of $J'$. $C(R)$ is nonmonotonic; the value of $R$
for which $C(R)$ is largest grows with the strength of the long-range
connection. The overall curve is lowered when $J'$ is increased.  As is
to be expected, once the ratio of $R$ to $J'$ becomes large enough that
the local fields dominate the system, the curves for $C(R)$ collapse.

The lowering of $C(R)$ with $J'$ means there are fewer small interevent
times for larger $J'$ values at a given disorder $R$.  However, despite
the shift in the distribution from smaller to larger interevent times,
$P(\Delta B)$ remains monotonically decreasing. These trends are present
as well in Fig.~\ref{fig:PDistribution}, where the distribution becomes
visibly flatter as $J'$ is increased.

\begin{figure}
\centering
\includegraphics[width =\columnwidth]{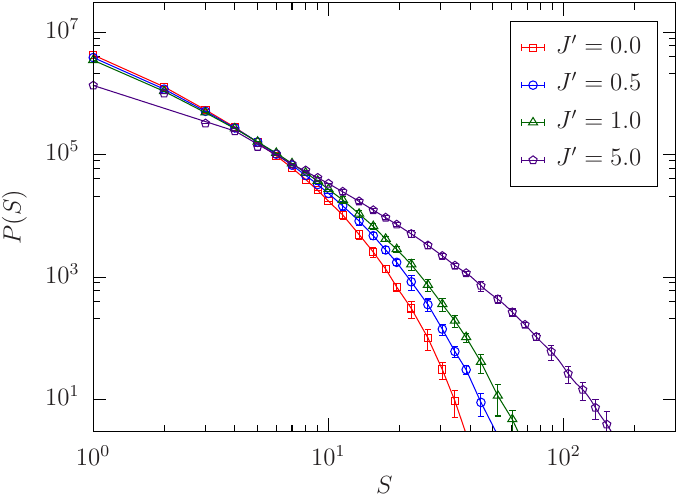}
\caption{
Distribution of avalanche sizes plotted for various $J'$ values for $N =
50^3$ and $R=7$. As $J'$ increases, there are larger avalanche
sizes, and the distribution approaches the critical region, as described
in Ref.~\cite{perkovic:95}. This is due to the correlation of lattice
sites. Note that two avalanches may occur at the same moment, which
leads to a greater amount of large avalanche sizes.}
\label{fig:SDistallJs}
\end{figure}

In order to explain this phenomenon, consider $J'=0$ and two avalanches
that occur separately in space and nearly simultaneously in time,
i.e.,~at very similar global field value $B$. In this moment, the two
events are uncorrelated, and they have a very small interevent time
$\Delta B$. If long-range connections are included, it is plausible that
the two avalanches might now be correlated and occur simultaneously as
$J'$ grows. The probability of small avalanches decreases and of large
ones increases, as seen in Fig.~\ref{fig:SDistallJs}. This also implies
that the frequency of small interevent times is reduced. The long-range
correlations do not ever result in a peak of $P(\Delta A)$ or $P(\Delta
F)$ which would indicate specific most probable {\em natural time
spacings} which predict when a large avalanche would be imminent.

Although $J'$ affects the quantitative value of $P(\Delta B)$, as well
as $P(\Delta F)$ and $P(\Delta A)$, it does not significantly change
their width.  In short, the conjecture that $J'$ might make avalanches constant in {\it natural time} seems to be false.

\section{\label{sec:RPM}Return point memory}

In the study of hysteresis loops, ``return point memory" is a central
concept \cite{katzgraber:02b,sethna:93,deutsch:04}. Instead of driving
the system to saturation, the external field $B$ is lowered from
infinity until some intermediate field $B_0$ is reached. At this point,
$B$ is raised to $B_1$ and then lowered back to $B_0$, which creates an
``internal" hysteresis loop. If the system exhibits return point memory
(RPM), then the state (that is, the magnetization) of the system is the
same at both instances of $B_0$.

It is natural to ask how the interevent times along an internal
hysteresis loop are distributed. It is known that both the RFIM and the
SKM exhibit RPM \cite{katzgraber:02b,pazmandi:99}. While the
distribution of interevent times in the SKM has been shown to be well
approximated by a Weibull distribution for sufficiently large $L_A$ (see
Fig.~\ref{fig:loglogBSK}), the RFIM interevent times are not as well
described by a Weibull distribution (see Fig.~\ref{fig:loglogBR}).
Therefore, we simulate the RFIM for both $R = 2.3$ and $R = 3$ on a
lattice size of $N = 200^3$ to study the interevent times for internal
hysteresis loops.  This helps identify whether RPM can help the RFIM
interevent distributions become Weibull-like, and the results can be
directly compared to Figs.~\ref{fig:weibullB} and \ref{fig:loglogBR}.

\begin{figure}
\centering
\includegraphics[width =\columnwidth]{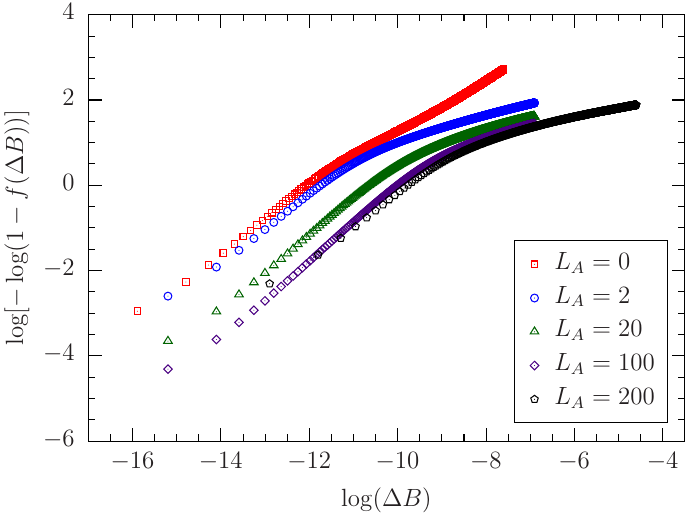}
\caption{
Cumulative distribution of the interevent times for the RFIM across an
internal hysteresis loop, which exhibits return point memory for $R=2.3$
and $N=200^3$. The data are more linear in a log-log plot for $L_A=0$
than $L_A=2$. Both $\Delta A$ and $\Delta F$ are similar to the results
for the whole hysteresis loop.
}
\label{fig:RPMloglogB2point3}
\end{figure}

\begin{figure}
\centering
\includegraphics[width =\columnwidth]{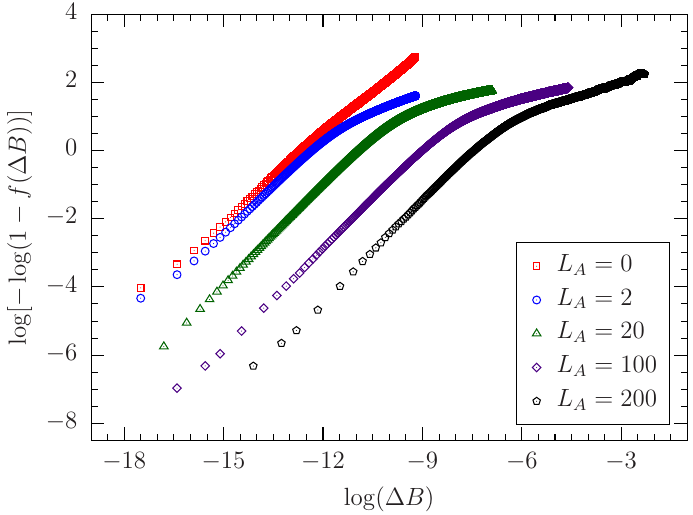}
\caption{
Cumulative distribution of the interevent times for the RFIM across an
internal hysteresis loop, which exhibits return point memory for $R=3$
and $N=200^3$. The data are Weibull distributed for $L_A=0$.
}
\label{fig:RPMloglogB3}
\end{figure}

The results for internal hysteresis loops are shown in
Figs.~\ref{fig:RPMloglogB2point3} and \ref{fig:RPMloglogB3}. The data
for $L_A=0$ approach a Weibull distribution even though the data are not
Weibull distributed for the smallest nonzero large avalanche. While the
data in a log-log plot are not exactly linear for $R=2.3$, there is
still a striking difference between $L_A=0$ and $L_A=2$. A more detailed
analysis on RPM is needed to fully understand the behavior of small
$L_A$. However, we expect the same trend to occur with alternate choices
of disorder and RPM turning points. Note that $\Delta A$ and $\Delta F$
are the same as the previous case when the system is driven to
saturation (see Fig.~\ref{fig:loglogFR}).


\section{\label{sec:conclusions}Conclusions}

We have evaluated the distribution of interevent times of the
three-dimensional random-field Ising, Sherrington-Kirkpatrick, and
Viana-Bray models.  Our motivation was twofold: First, to extend the
analytic results for this distribution---which have been obtained in one
space dimension \cite{nampoothiri:17}--- to higher space dimensions,
providing complementary numerical results to the well-studied
distributions of the avalanche amplitudes.  Second, to explore the idea
of natural time to study if the distribution is more sharply peaked when
measured by counting the number of small avalanches rather than the
change in field itself.

Our conclusions regarding the first point are summarized in
Figs.~\ref{fig:deltaH} and \ref{fig:SKdist}, which provide explicit
forms for $P(\Delta B)$ for the different models.  A central feature of
our results is the scaling relation, $P(N,\Delta B) \sim \tilde
P(N\Delta B)$ obeyed by both the RFIM and VBM, whose validity we trace
to a hysteresis loop width which is nearly independent of lattice size
$N$ (Fig.~\ref{fig:RFhys}). In contrast, the SKM hysteresis loop width
changes significantly with $N$ (Fig.~\ref{fig:SKhys}), and $P(N,\Delta
B)$ does not scale.  We also observe a pseudogap exponent $\theta
\approx 0$, which is the same as in the one-dimensional case
\cite{nampoothiri:17}. By examining the dependence of the interevent
distribution on the disorder, one finds nonmonotonic behavior where
$C(R)$ [Eq.~\eqref{eq:CR}] peaks at $\sim 3.7$, similar to the
one-dimensional case.

Regarding the second point, we have added a large avalanche threshold,
similar to the large earthquake threshold used in geophysics
\cite{varotsos:06}. This leads to a distribution of interevent times
for several methods of counting time. Counting the number of individual
spin flips, counting the small avalanches, and the original measurement
in terms of the change in magnetic field.  We see no evidence for a
sharpening of the interevent time distribution function which would be
a confirmation that large events occur at a specific $\Delta A$ or
$\Delta F$.  The clock-time fit to a Weibull distribution, however, is
improved by the introduction of an event threshold $L_A$
(Fig.~\ref{fig:loglogBSK}) in the SKM. The distribution of the natural
time $\Delta A$ can also be fit to a Weibull distribution in the SKM
(Fig.~\ref{fig:loglogASK}).

Finally, we applied the same analysis to interevent distributions for a
model system with added small-world bonds, i.e., bonds between random
pairs of lattice sites. As the strength of these small-world bonds is
increased, there are fewer small interevent times and fewer small
avalanches. This is due to the increased correlation of lattice sites.
As the correlation increases, avalanches coalesce into large avalanches,
which reduces the number of small interevent times and small avalanche
sizes. In some sense, the strength of the long-range bonds could be
thought of as a tuning parameter between the distribution of avalanche
sizes and interevent times.

By adding the long-range bonds to the lattice, the interevent
distributions of large avalanches do not change.  There is one main
difference when long-range bonds are added. Namely, the value of the
critical disorder increases when the strength of the long-range
interactions are increased. As long as the disorder relative to the
critical region is the same, the statistics of large avalanche
interevent times is the same for any strength of the long-range
interactions. Thus this modification of the model to make natural time a
more effective clock is seen not to be effective.

There are several possible objectives to a quantitative evaluation of
interevent times.  One goal is a determination of their distribution
function.  In such an investigation, it is possible that alternate
definitions of time, $\Delta B$, $\Delta A$, or $\Delta F$ in the work
reported here, might lead to more simple or well understood
distributions.  We have shown that in the SKM, the use of a finite event
threshold and $\Delta A$ simplifies the nature of $P(\Delta A)$ to a
Weibull distribution.

A second goal concerns the {\it prediction} of the next (large) event.
That requires not only finding the distribution function, but also,
through the use of an appropriate redefinition of time, acquiring
a distribution function which is sharply peaked, so the separation
between events is known.  This is, obviously, a holy grail for
earthquake prediction.  We have not succeeded in finding such a
transformation for interacting spin models.  Nevertheless, we suggest
that further exploring the idea within simple models might be a useful,
more controllable, complement to analysis of observational data.

\begin{acknowledgments}

We thank Alexis Giguere, Molly Luginbuhl, John Rundle, and Don Turcotte
for discussions on geophysics, Mario D'Andrea for initial coding,
discussion, and writing, Juan Carlos Andresen for sharing base software, and Texas
A\&M University for access to their HPC resources. H.G.K.~would like
to thank Mezcal Reposado for motivation during the final stages of the
manuscript.  The work of R.S.~and J.F.~was supported by the Department
of Energy, Grant No.~DE-SC0014671.  H.G.K.~and A.B.~acknowledge
support from the NSF (Grant No.~DMR-1151387). H.G.K.'s research is
based upon work supported by the Office of the Director of National
Intelligence (ODNI), Intelligence Advanced Research Projects Activity
(IARPA), via Interagency Umbrella Agreement No.~IA1-1198. The views and
conclusions contained herein are those of the authors and should not be
interpreted as necessarily representing the official policies or
endorsements, either expressed or implied, of the ODNI, IARPA, or the
U.S.~Government.  The U.S.~Government is authorized to reproduce and
distribute reprints for Governmental purposes notwithstanding any
copyright annotation thereon.

\end{acknowledgments}

\bibliography{refs}

\end{document}